\documentstyle{article}
\def\1ad{\mbox{\normalsize $^1$}}
\def\2ad{\mbox{\normalsize $^2$}}
\def\3ad{\mbox{\normalsize $^3$}}
\def\4ad{\mbox{\normalsize $^4$}}
\def\5ad{\mbox{\normalsize $^5$}}
\def\6ad{\mbox{\normalsize $^6$}}
\def\7ad{\mbox{\normalsize $^7$}}
\def\8ad{\mbox{\normalsize $^8$}}
\def\makefront{\vspace*{1cm}\begin{center}
\def\newtitleline{\\ \vskip 5pt}
{\Large\bf\titleline}\\
\vskip 1truecm
{\large\bf\authors}\\
\vskip 5truemm
\addresses
\end{center}
\vskip 1truecm
{\bf Abstract:}
\abstracttext
\vskip 1truecm}
\newcommand{\eqn}[1]{(\ref{#1})}

\newsavebox{\uuunit}
\sbox{\uuunit}
    {\setlength{\unitlength}{0.825em}
     \begin{picture}(0.6,0.7)
        \thinlines
        \put(0,0){\line(1,0){0.5}}
        \put(0.15,0){\line(0,1){0.7}}
        \put(0.35,0){\line(0,1){0.8}}
       \multiput(0.3,0.8)(-0.04,-0.02){12}{\rule{0.5pt}{0.5pt}}
     \end {picture}}

\def\IP{\relax{\rm I\kern-.18em P}}

\font\cmss=cmss10 \font\cmsss=cmss10 at 7pt

\def\inbar{\vrule height1.5ex width.4pt depth0pt}
\def\IC{\relax\,\hbox{$\inbar\kern-.3em{\rm C}$}}
\def\IG{\relax\,\hbox{$\inbar\kern-.3em{\rm G}$}}
\def\IB{\relax{\rm I\kern-.18em B}}
\def\ID{\relax{\rm I\kern-.18em D}}
\def\IL{\relax{\rm I\kern-.18em L}}
\def\IF{\relax{\rm I\kern-.18em F}}
\def\IH{\relax{\rm I\kern-.18em H}}
\def\II{\relax{\rm I\kern-.17em I}}
\def\IN{\relax{\rm I\kern-.18em N}}
\def\IP{\relax{\rm I\kern-.18em P}}
\def\IQ{\relax\,\hbox{$\inbar\kern-.3em{\rm Q}$}}
\def\bfzero{\relax\,\hbox{$\inbar\kern-.3em{\rm 0}$}}
\def\IK{\relax{\rm I\kern-.18em K}}
\def\IG{\relax\,\hbox{$\inbar\kern-.3em{\rm G}$}}
 \font\cmss=cmss10 \font\cmsss=cmss10 at 7pt
\def\IR{\relax{\rm I\kern-.18em R}}
\def\ZZ{\relax\ifmmode\mathchoice
{\hbox{\cmss Z\kern-.4em Z}}{\hbox{\cmss Z\kern-.4em Z}}
{\lower.9pt\hbox{\cmsss Z\kern-.4em Z}}
{\lower1.2pt\hbox{\cmsss Z\kern-.4em Z}}\else{\cmss Z\kern-.4em
Z}\fi}
\def\bfone{\relax{\rm 1\kern-.35em 1}}

\def\bfone{\relax{\rm 1\kern-.35em 1}}
\font\cmss=cmss10 \font\cmsss=cmss10 at 7pt

\newcommand{\be}{\begin{equation}}
\newcommand{\ee}{\end{equation}}
\newcommand{\bea}{\begin{eqnarray}}
\newcommand{\eea}{\end{eqnarray}}
  
\def\tilde{\widetilde}
\def\bar{\overline}

\def\Coe#1.#2.{{#1\over #2}}

\def\coe#1.#2.{\relax{\textstyle {#1 \over #2}}\displaystyle}

\def\notin{\hbox{{$\in$}\kern-.51em\hbox{/}}}

\def\IE{\relax{{\rm I\kern-.18em E}}}

\def\IGam{\relax{{\rm I}\kern-.18em \Gamma}}

\begin{document}
\def\titleline{
Solvable Lie Algebras, BPS Black Holes
\newtitleline
and Supergravity Gaugings
}
\def\authors{Pietro Fr\'e}
\def\addresses{
Dipartimento di Fisica Teorica, \\
Universit\'a di Torino
Via P. Giuria 1, I-10125 TORINO\\
Italy
}
{\it Talk given at EEC TMR contract meeting in Neuchatel, Gauge
Theories, SUSY, Unification, September 1997}
\def\abstracttext{
In this talk I review both accomplished results and work
in progress on the use of solvable Lie algebras as an intrinsic
algebraic characterization of the scalar field sector of M--theory low
energy effective lagrangians. In particular I review the application
of these techniques in obtaining the most general form of BPS
black hole solutions.
}
\makefront
\section{Introduction }
In this talk I review a line of research \cite{noi1},\cite{noi2},\cite{noi3}
I have recently pursued in
collaboration with Riccardo D'Auria, Sergio Ferrara and our Ph.D
students Laura Andrianopoli and Mario Trigiante, whose contribution
to the development of the entire project has been essential.
\par
The main idea underlying this investigation  originates from some
results I had previously obtained in collaboration with Luciano
Girardello, Igor Pesando and Mario Trigiante. In \cite{lucianoi},
extending work of Girardello, Ferrara and Porrati, we discovered that
$N=2$ supersymmetry can be spontaneously broken to $N=1$ when the
following conditions are met:
\begin{itemize}
\item The scalar manifold of supergravity, which is generically given
by the direct product ${\cal SK} \otimes {\cal QK}$ of a special K\"ahler
manifold with a quaternionic one, is a {\it homogeneous non--compact
coset manifold} ${\cal G}/{\cal H}$
\item Some {\it translational abelian} symmetries of ${\cal G}/{\cal H}$
are gauged.
\end{itemize}
The basic ingredient in deriving the above result is the Alekseevskian
description \cite{alex} of the scalar manifold ${\cal SK} \otimes {\cal QK}$ in
terms of {\it solvable} Lie algebras, a K\"ahler algebra ${\cal K}$ for the vector
multiplet sector ${\cal SK}$ and a quaternionic algebra ${\cal Q}$
for the hypermultiplet sector ${\cal QK}$. By means of this
description the homogeneous non compact coset manifold ${\cal G}/{\cal H}$
is identified with the {\it solvable group manifold} $\exp[Solv]$
where
\begin{equation}
  Solv  = {\cal K} \, \oplus  \, {\cal Q}  \label{k+q}
\end{equation}
and the translational symmetries responsible for the supersymmetry
breaking are identified with suitable abelian subalgebras
\begin{equation}
{\cal T} \, \subset \, Solv     \label{tinsolv}
\end{equation}
An obvious observation that easily occurs once such a perspective is
adopted is the following one: for all extended supergravities with
$N \ge 3$ the scalar manifold is a   homogeneous non--compact
coset manifold  ${\cal G}/{\cal H}$. Hence it is very tempting to extend
the {\it solvable Lie algebra approach} to such supergravity theories, in
particular to the maximal extended ones in all dimensions $4 \le D
\le 10$. This is what was done in the series of three papers
\cite{noi1,noi2,noi3}.
\par
\section{R-R and NS-NS scalars}
Relying on a well established mathematical theory which is available
in standard textbooks (for instance
\cite{helgason}), every non compact homogeneous space ${\cal G}/{\cal H}$
is indeed a solvable group manifold and its generating solvable Lie
algebra $Solv \left( {\cal G}/{\cal H} \right)$ can be constructed
utilizing roots and Dynkin diagram techniques. This fact
offers the so far underestimated possibility of introducing an
intrinsic algebraic characterization of the supergravity scalars. In
relation with string theory this yields
a group--theoretical definition of Ramond  and Neveu--Schwarz scalars.
It goes as follows. The same supergravity lagrangian admits different
interpretations as low energy theory of different superstrings
related by duality transformations or of M--theory. The
identification of the Ramond and Neveu Schwarz sectors is different in the
different interpretations. Algebraically this corresponds to
inequivalent decompositions of the solvable Lie algebra
$Solv \left( {\cal G}/{\cal H} \right)$ with respect to different subalgebras.
Each string theory admits a $T$--duality and an $S$--duality group
whose product $S \otimes   T$ constitutes a subgroup of the
$U$--duality group, namely of the isometry group $U \equiv {\cal G}$
of the homogenoeus scalar manifold ${\cal G}/{\cal H}$. Physically
$S$ is a non perturbative symmetry acting on the {\it dilaton} while $T$ is
a perturbative symmetry acting on the "{\it radii}" of the
compactification. There exist also two compact subgroups ${\cal H}_S
\subset S$ and ${\cal H}_T \subset T$ whose product ${\cal H}_S \otimes
{\cal H}_T \,\subset  \, H$ is contained in the maximal compact
subgroup ${\cal H} \subset U$ such that we can write:
\begin{equation}
Solv \left( {\cal U}/{\cal H} \right) = Solv \left( { S}/{\cal H}_S \right)
\, \oplus  \,  Solv \left( { T}/{\cal H}_T \right) \,
\oplus \, {\cal W}
\end{equation}
the three addends being all subalgebras of
$Solv \left( {\cal U}/{\cal H} \right)$. The first two addends
constitute the Neveu Schwarz sector while the last subalgebra ${\cal W}$
which is not only solvable but also {\it nilpotent} constitutes the
Ramond sector  relative to the chosen superstring interpretation.
\par
An example of this way of reasoning is provided by maximal
supergravities in $D=10-r$ dimensions. For such lagrangians the
scalar sector is given by ${\cal M}_{scalar} = E_{r+1(r+1)}/{\cal H}_{r+1}$
where the group $E_{r+1(r+1)}$ is obtained exponentiating
the maximally non compact real form of the exceptional rank $r+1$ Lie
algebra $E_{r+1}$ and ${\cal H}_{r+1}$ is the corresponding maximal
compact subgroup. If we interpret supergravity as the low energy
theory of Type IIA superstring compactified on a torus $T^r$, then
the appropriate $S$-duality group is $O(1,1)$ and the appropriate
$T$--duality group is $SO(r,r)$. Correspondingly we obtain the
decomposition:
\begin{equation}
Solv \left(  E_{r+1(r+1)}/ {\cal H}_{r+1} \right )
= O(1,1)
\, \oplus  \,  Solv \left(   \frac {SO(r,r)}{SO(r)\times SO(r)  }  \right) \,
\oplus \, {\cal W}_{r+1}
\label{equata}
\end{equation}
where the Ramond subalgebra ${\cal W}_{r+1}\equiv  spin[r,r]  $
is nothing else but the
chiral spinor representation of   $SO(r,r)$.
In the four dimensional case $r=6$ equation
\eqn{equata} takes the exceptional form:
\begin{equation}
Solv \left(  E_{7(7)}/ SU(8) \right )
= Solv \left(  SL(2,R)/ O(2) \right )
\, \oplus  \,  Solv \left(   \frac {SO(6,6)}{SO(6)\times SO(6)  }  \right) \,
\oplus \, {\cal W}_{7}
\label{equadue}
\end{equation}
The 38 Neveu Schwarz scalars are given by the first two addends in
\eqn{equadue}, while the 32 Ramond scalars in the algebra ${\cal W}_{7}$
transform in the spinor representation of $SO(6,6)$ as in all the other cases.
\par
Alternatively we can interpret maximal supergravity in $D=10-r$ as
the compactification on a torus $T^r$ of Type IIB superstring. In
this case the ST--duality group is different. We just have:
\begin{equation}
S \, \otimes  \, T \, = \, O(1,1) \, \otimes \, GL(r) \label{equatre}
\end{equation}
Correspondingly we write the solvable Lie algebra decomposition:
\begin{equation}
Solv \left(  E_{r+1(r+1)}/ {\cal H}_{r+1} \right )
= O(1,1)
\, \oplus  \,  Solv \left(   \frac {GL(r)}{SO(r) }  \right) \,
\oplus \, {\tilde {\cal W}}_{r+1}
\label{equaquat}
\end{equation}
where ${\tilde {\cal W}}_{r+1}$ is the new algebra of Ramond scalars
with respect to the Type IIB interpretation. Actually, as it is
well known, Type IIB theory already admits an $SL(2,R)$ U--duality symmetry
in ten dimensions that mixes Ramond and Neveu Schwarz states. The proper
S--duality group $O(1,1)$ is just a maximal subgroup of such $SL(2,R)$.
Correspondingly eq. \eqn{equaquat} can be restated as:
\begin{equation}
Solv_{r+1}\, \equiv \,Solv \left(  E_{r+1(r+1)}/ {\cal H}_{r+1} \right )
= Solv \left(  SL(2,R)/ O(2) \right )
\, \oplus  \,  Solv \left(   \frac {GL(r)}{SO(r) }  \right) \,
\oplus \, {\bar {\cal W}}_{r+1}
\label{equacinq}
\end{equation}
Finally a third decomposition of the same solvable Lie algebra can be
written if the same supergravity lagrangian is intepreted as
compactification on a torus $T^7$ of M--theory. For the details on
this and other decompositions of the scalar sector
that keep track of the sequential compactifications on multiple torii
we refer the reader to the original papers \cite{noi1,noi2}.
\section{BPS black holes}
Another interesting application of the solvable Lie algebra
parametrization of the scalar sector is provided by the systematic
construction of completely general BPS black hole or BPS black brane solutions.
In this  problem  a fundamental role is played by the
scalar evolution from arbitrary values at infinity
to fixed values at the horizon. Such an evolution is best understood
and dealt with when the scalars are algebraically characterized as
generators of a solvable algebra. In ref.\cite{noi3} we considered
the $N=8$,$D=4$ case and we solved the problem of writing the most
general BPS black hole solution that preserves $1/8$ of the original
supersymmetries. We are presently pursuing the solution of the same
problem in the case where the preserved supersymmetries are either
$1/2$ or $1/4$ \cite{nuovonoi}. Let us briefly illustrate these three
cases from our viewpoint.
\par
 The $D=4$ supersymmetry algebra with
$N=8$  supersymmetry charges  can be written in the following
form:
\begin{eqnarray}
&\left\{ {\bar Q}_{aI \vert \alpha }\, , \,{\bar Q}_{bJ \vert \beta}
\right\}\, = \,  {\rm i} \left( C \, \gamma^\mu \right)_{\alpha \beta} \,
P_\mu \, \delta_{ab} \, \delta_{IJ} \, - \, C_{\alpha \beta} \,
\epsilon_{ab} \, \times \, \ZZ_{IJ}& \nonumber\\
&\left( a,b = 1,2 \qquad ; \qquad I,J=1,\dots, 4 \right)&
\label{susyeven}
\end{eqnarray}
where the SUSY charges ${\bar Q}_{aI}\equiv Q_{aI}^\dagger \gamma_0=
Q^T_{ai} \, C$ are Majorana spinors, $C$ is the charge conjugation
matrix, $P_\mu$ is the 4--momentum operator, $\epsilon_{ab}$ is the
two--dimensional Levi Civita symbol and the symmetric tensor
$\ZZ_{IJ}=\ZZ_{JI}$ is the central charge operator. It
can always be diagonalized $\ZZ_{IJ}=\delta_{IJ} \, Z_J$ and
its $4$ eigenvalues $Z_J$ are the central charges.
\par
Consider the reduced supercharges:
\begin{equation}
{\bar S}^{\pm}_{aI \vert \alpha }=\frac{1}{2} \,
\left( {\bar Q}_{aI} \gamma _0 \pm \mbox{i} \,
\epsilon_{ab} \,  {\bar Q}_{bI}\,
\right)_\alpha
\label{redchar}
\end{equation}
They can be regarded as the result of applying
a projection operator to the supersymmetry
charges: $ {\bar S}^{\pm}_{aI}  =  {\bar Q}_{bI} \, \IP^\pm_{ba} $,
where $ \IP^\pm_{ba} = \frac{1}{2}\, \left({\bf 1}\delta_{ba} \pm \mbox{i}
\epsilon_{ba} \gamma_0 \right)$. In the rest frame where the four momentum
is $P_\mu$ =$(M,0,0,0)$, we obtain the algebra:
$
\left\{ {\bar S}^{\pm}_{aI}  \, , \, {\bar S}^{\pm}_{bJ} \right\} =
\pm \epsilon_{ac}\, C \, \IP^\pm_{cb} \, \left( M \mp Z_I \right)\,
\delta_{IJ}
$ and the BPS states that saturate the bounds $
\left( M\pm Z_I \right) \, \vert \mbox{BPS state,} i\rangle = 0
$ are those which are annihilated by the corresponding reduced supercharges:
\begin{equation}
{\bar S}^{\pm}_{aI}   \, \vert \mbox{BPS state,} i\rangle = 0
\label{susinvbps}
\end{equation}
Eq.\eqn{susinvbps} defines {\sl short multiplet
representations} of the original algebra \eqn{susyeven} in the
following sense: one constructs a linear representation of \eqn{susyeven}
where all states are identically
annihilated by the operators ${\bar S}^{\pm}_{aI}$ for $I=1,\dots,n_{max}$.
If $n_{max}=1$ we have the minimum shortening, if $n_{max}=4$ we
have the maximum shortening. On the other hand eq.\eqn{susinvbps}
can be translated into first order differential equations on the
bosonic fields of supergravity  whose common solutions with the ordinary
field  equations are the BPS saturated black hole configurations. In
the case of maximum shortening $n_{max}=4$ the black hole preserves
$1/2$ supersymmetry, in the case of intermediate shortening
$n_{max}=2$  it preserves $1/4$, while in the case of minimum
shortening it preserves  $1/8$.
\subsection{The Killing spinor equation and its covariance group }
In order to translate eq.\eqn{susinvbps} into first order differential equations on the
bosonic fields of supergravity we consider
a configuration where all the fermionic fields are zero and we set
to zero the fermionic SUSY rules appropriate to such a background
\begin{equation}
0=\delta \mbox{fermions} = \mbox{SUSY rule} \left( \mbox{bosons},\epsilon_{Ai} \right)
\label{fermboserule}
\end{equation}
and to a SUSY parameter that satisfies the following conditions:
\begin{equation}
\begin{array}{rclcl}
\xi^\mu \, \gamma_\mu \,\epsilon_{aI} &=& \mbox{i}\, \varepsilon_{ab}
\,  \epsilon^{bI}   & ; &   i=1,\dots,n_{max}\\
\epsilon_{aI} &=& 0  &;&   i > n_{max} \\
\end{array}
\label{kilspieq}
\end{equation}
Here $\xi^\mu$ is a time--like Killing vector for the space--time metric and
$ \epsilon _{aI}, \epsilon^{aI}$ denote the two chiral projections of
a single Majorana spinor: $ \gamma _5 \, \epsilon _{aI} \, = \, \epsilon _{aI} $ ,
$ \gamma _5 \, \epsilon ^{aI} \, = - \epsilon ^{aI} $
We name eq.\eqn{fermboserule} the {\sl Killing spinor equation}
and the investigation of its
group--theoretical structure was our main goal in ref \cite{noi3}.
There we restricted our attention to the case $n_{max}=1$: we
are presently considering the other two possibilities \cite{nuovonoi}.
In all three cases eq.\eqn{fermboserule} has two features which we want to stress
as main motivations for the developments we have pursued:
\begin{enumerate}
\item{It requires an efficient parametrization of the scalar field
sector}
\item{It breaks the original $SU(8)$ automorphism group of the
supersymmetry algebra to the subgroup $Usp(2 \,n_{max})\times SU(8-2\,n_{max})\times U(1)$}
\end{enumerate}
The first feature is the reason why the use of the rank $7$ solvable Lie
algebra $Solv_7$ associated with $E_{7(7)}/SU(8)$ is of great help in this problem.
The second feature is the
reason why the solvable Lie algebra $Solv_7$ has to be decomposed in
a way appropriate to the decomposition of the isotropy group
$SU(8)$
with respect to the subgroup $Usp(2\,n_{max})\times SU(8-2\,n_{max})\times U(1)$.
\par
This decomposition of the solvable Lie algebra is a close relative of
the decomposition of $N=8$ supergravity into multiplets of the lower
supersymmetry $N^\prime = 2 \, n_{max}$. This is easily understood by
recalling that close to the horizon of the black hole one doubles the supersymmetries
holding in the bulk of the solution. Hence the near horizon
supersymmetry is precisely $N^\prime = 2 \, n_{max}$ and the black
solution can be interpreted as a soliton that interpolates between
{\it ungauged} $N=8$ supergravity at infinity and some form of {\it
gauged} $N^\prime$ supergravity at the horizon. The reason why we
stress that the horizon theory is gauged is that its geometry is an
anti de Sitter geometry.
\par
Let us now study the explicit structure of the three cases at hand.
\subsubsection{The $1/2$ SUSY case}
Here we have $n_{max} = 4$ and correspondingly the covariance subgroup
of the Killing spinor equation is $Usp(8)\, \subset \, SU(8)$. Indeed
 condition \eqn{kilspieq} can be rewritten as follows:
 \begin{equation}
\xi^\mu \, \gamma_\mu \,\epsilon_{A}  =  \mbox{i}\, {\cal C}_{AB}
\,  \epsilon^{B} \quad ; \quad A,B=1,\dots ,8
\label{urcunmez}
\end{equation}
where ${\cal C}_{AB}= - {\cal C}_{BA}$ denotes an $ 8 \times 8$
antisymmetric matrix satisfying ${\cal C}^2 = -\bfone$. The group
$Usp(8)$ is the subgroup of unimodular, unitary $ 8 \times 8$
matrices that are also symplectic, namely that preserve the matrix
${\cal C}$.
\par
We are accordingly lead to decompose the solvable Lie algebra as written below:
\begin{eqnarray}
Solv_7 & =& Solv_6 \, \oplus \, O(1,1) \, \oplus \ID_6
\label{decomp1a}\\
70 & = & 42 \, + \, 1 \, + \, 27
\label{decomp1b}
\end{eqnarray}
where, following the notation established in \eqn{equacinq}:
\begin{eqnarray}
Solv_7 & \equiv & Solv \left(\frac{E_{7(7)}}{SU(8)}\right)
\nonumber\\
Solv_6 & \equiv & Solv \left(\frac{E_{6(6)}}{Usp(8)}\right)
\nonumber\\
\mbox{dim}\, Solv_7 & = & 70 \quad ; \quad \mbox{rank}\, Solv_7 \, =
\, 7 \nonumber\\
\mbox{dim}\, Solv_6 & = & 42 \quad ; \quad \mbox{rank}\, Solv_6 \, =
\, 6 \nonumber\\
\label{defsolv7}
\end{eqnarray}
In eq.\eqn{decomp1a}
 $Solv_6$ is the solvable Lie algebra that describes the scalar
sector of $D=5$, $N=8$ supergravity, while the $27$--dimensional
abelian ideal $\ID_6$ corresponds to those $D=4$ scalars that
originate from the $27$--vectors of supergravity one--dimension
above \cite{noi2}.
Eq.\eqn{decomp1b} corresponds also to the decomposition of the ${\bf
70}$ irreducible representation of $SU(8)$ into irreducible
representations of $Usp(8)$. Indeed we have:
\begin{equation}
{\bf 70} \, \stackrel{Usp(8)}{\longrightarrow} \, {\bf 42} \, \oplus
\, {\bf 1} \, \oplus \, {\bf 27}
\label{uspdecompo1}
\end{equation}
In order to  single out the content of the first order Killing
spinor equations we need to decompose them into irreducible $Usp(8)$ representations.
This is easily done. The gravitino equation is an ${\bf 8}$ of $SU(8)$ that
remains irreducible under $Usp(8)$ reduction. On the other hand the dilatino
equation is a ${\bf 56}$ of $SU(8)$ that reduces as follows:
\begin{equation}
{\bf 56} \, \stackrel{Usp(8)}{\longrightarrow} \, {\bf 48} \, \oplus
\, {\bf 8}
\label{uspdecompo2}
\end{equation}
Hence altogether we have the following three constraints  ${\bf 8}$,
${\bf 8}^\prime$ , ${\bf 48}$ on the three subalgebras of scalar
fields ${\bf 42}$, ${\bf 1}$ and ${\bf 27}$. Working out
the consequences of these constraints and deciding which scalars are
set to constants and which are insteading evolving is work in
progress \cite{nuovonoi}.
\subsubsection{The $1/4$ SUSY case}
Here we have $n_{max} = 2$ and correspondingly the covariance subgroup
of the Killing spinor equation is $Usp(4)\,\times \, SU(4) \,\times \, U(1)
\subset \, SU(8)$. Indeed
 condition \eqn{kilspieq} can be rewritten as follows:
 \begin{eqnarray}
\xi^\mu \, \gamma_\mu \,\epsilon_{a} & =&  \mbox{i}\, {\cal C}_{ab}
\,  \epsilon^{b} \quad ; \quad a,b=1,\dots ,4 \nonumber\\
\epsilon_{X} & =& 0; \quad X=5,\dots ,8 \nonumber\\
\label{urcunquart}
\end{eqnarray}
where ${\cal C}_{ab}= - {\cal C}_{ba}$ denotes a  $ 4 \times 4$
antisymmetric matrix satisfying ${\cal C}^2 = -\bfone$. The group
$Usp(4)$ is the subgroup of unimodular, unitary $ 4 \times 4$
matrices that are also symplectic, namely that preserve the matrix
${\cal C}$.
\par
We are accordingly lead to decompose the solvable Lie algebra as follows.
Naming:
\begin{eqnarray}
Solv_S & \equiv & Solv \left(\frac{SL(2,R)}{U(1)}\right)
\nonumber\\
Solv_T & \equiv & Solv \left(\frac{SO(6,6)}{SO(6) \times SO(6)}\right)
\nonumber\\
\mbox{dim}\, Solv_S & = & 2 \quad ; \quad \mbox{rank}\, Solv_S \, =
\, 1 \nonumber\\
\mbox{dim}\, Solv_T & = & 36 \quad ; \quad \mbox{rank}\, Solv_T \, =
\, 6 \nonumber\\
\label{defsolst}
\end{eqnarray}
we can write:
\begin{eqnarray}
Solv_7 & =& Solv_S \, \oplus \, Solv_T \, \oplus \, {\cal W}_{7}
\label{stdecomp1a}\\
70 & = & 2\, + \, 36 \, + \, 32
\label{stdecomp1b}
\end{eqnarray}
which is nothing else but \eqn{equadue}. Indeed
 the solvable Lie algebras $Solv_S$ and $Solv_T$ describe the dilaton--axion
sector and the six torus moduli, respectively, in the interpretation
of $N=8$ supergravity as the compactification of Type IIA theory on a six--torus
$T^6$ \cite{noi2}. The rank zero abelian subalgebra ${\cal W}_{7}$
is instead composed by the Ramond-Ramond scalars as we have already explained.
\par
Introducing the decomposition \eqn{stdecomp1a}, \eqn{stdecomp1b} we have
succeeded in singling out a holonomy subgroup $SU(4) \, \times \,
SU(4) \, \times \, U(1) \, \subset \, SU(8)$. Indeed we have
$SO(6) \, \equiv \, SU(4)$. This is a step forward but it is not yet
the end of the story since we actually need a subgroup $Usp(4) \,
\times \, SU(4)$. This means that we must further decompose the
solvable Lie algebra $Solv_T$. This latter is the manifold of the
scalar fields associated with vector multiplets in an $N=4$
decomposition of the $N=8$ theory. Indeed the decomposition
\eqn{stdecomp1a} with respect to the S--T--duality subalgebra is the
appropriate decomposition of the scalar sector according to $N=4$
multiplets. The further decomposition we need is the following:
\begin{eqnarray}
Solv_T &=& Solv_{T5} \oplus Solv_{T1}  \nonumber\\
Solv_{T5}& \equiv &Solv \left( \frac{SO(5,6)}{SO(5) \times
SO(6)}\right)\nonumber  \\
Solv_{T1}& \equiv &Solv \left( \frac{SO(1,6)}{
SO(6)}\right)\nonumber  \\
\label{stspilt}
\end{eqnarray}
where we rely on the isomorphism $Usp(4) \equiv SO(5)$. Hence,
altogether we can write:
\begin{eqnarray}
Solv_7 &=& Solv_S \, \oplus \, Solv_{T5} \, \oplus \, Solv_{T1} \,
\oplus \, {\cal W}_{32} \nonumber\\
70 & = & 2 \, + \, 30 \, + \, 6 \, + \, 32
\label{sputast}
\end{eqnarray}
which exactly corresponds to the decomposition of the ${\bf 70}$
irreducible representation of $SU(8)$ into irreducible
representations of $Usp(4) \, \times \, SU(4) \, \times \, U(1)$.
\begin{equation}
{\bf 70} \, \stackrel{Usp(4) \, \times \, SU(4) \, \times \, U(1)}{\longrightarrow}
\, \left( {\bf 1}+{\bar {\bf 1}},{\bf 1},{\bf 1}\right) \, \oplus
\, \left( {\bf 1} ,{\bf 5},{\bf 6}\right) \, \oplus \,
\left( {\bf 1} ,{\bf 1},{\bf 6}\right)
\label{uspdecompo3}
\end{equation}
Just as in the previous case we should now
single out the content of the first order Killing
spinor equations by decomposing  them into irreducible
$Usp(4) \, \times \, SU(4) \, \times \, U(1)$ representations. This
is also work in progress \cite{nuovonoi}.
\subsubsection{The $1/8$ SUSY case}
Here we have $n_{max} = 1$ and
$Solv_7$ must be decomposed  according to the decomposition
of the isotropy subgroup: $SU(8) \longrightarrow SU(2)\times U(6)$. We
showed in \cite{noi3}  that the corresponding decomposition of the solvable
Lie algebra is the following one:
\begin{equation}
Solv_7   =  Solv_3 \, \oplus \, Solv_4
\label{7in3p4}
\end{equation}
\begin{equation}
\begin{array}{rclrcl}
Solv_3 & \equiv & Solv \left( SO^\star(12)/U(6) \right) & Solv_4 &
\equiv & Solv \left( E_{6(4)}/SU(2)\times SU(6) \right) \\
\mbox{rank }\, Solv_3 & = & 3 & \mbox{rank }\,Solv_4 &
= & 4 \\
\mbox{dim }\, Solv_3 & = & 30 & \mbox{dim }\,Solv_4 &
= & 40 \\
\end{array}
\label{3and4defi}
\end{equation}
The rank three  Lie algebra $Solv_3$ defined above describes the
thirty dimensional scalar sector of $N=6$ supergravity, while the rank four
solvable Lie algebra $Solv_4$ contains the remaining forty scalars
belonging to $N=6$ spin $3/2$ multiplets. It should be noted
that, individually, both manifolds $ \exp \left[ Solv_3 \right]$ and
$ \exp \left[ Solv_4 \right]$ have also an $N=2$ interpretation since we have:
\begin{eqnarray}
\exp \left[ Solv_3 \right] & =& \mbox{homogeneous special K\"ahler}
\nonumber \\
\exp \left[ Solv_4 \right] & =& \mbox{homogeneous quaternionic}
\label{pincpal}
\end{eqnarray}
so that the first manifold can describe the interaction of
$15$ vector multiplets, while the second can describe the interaction
of $10$ hypermultiplets. Indeed if we decompose the $N=8$ graviton
multiplet in $N=2$ representations we find:
\begin{equation}
\mbox{N=8} \, \mbox{\bf spin 2}  \,\stackrel{N=2}{\longrightarrow}\,
 \mbox{\bf spin 2} + 6 \times \mbox{\bf spin 3/2} + 15 \times \mbox{\bf vect. mult.}
 +
 10 \times \mbox{\bf hypermult.}
 \label{n8n2decompo}
\end{equation}
Introducing the decomposition \eqn{7in3p4}
we found in \cite{noi3} that the $40$ scalars belonging to $Solv_4$ are constants
independent of the radial variable $r$. Only the $30$ scalars in the
K\"ahler algebra $Solv_3$ can be radial dependent. Infact their
radial dependence is governed by a first order differential equation
that can be extracted from a suitable component of the Killing spinor
equation. More precisely we obtained the following result.
Up to U--duality transformations the most general $N=8$ black--hole
is actually an $N=2$ black--hole corresponding to a very
specific choice of the special K\"ahler manifold, namely $ \exp[ Solv_3 ]$
as in eq.\eqn{pincpal},\eqn{3and4defi}. Furthermore up to the duality
rotations of $SO^\star(12)$ this general solution is actually
determined by the so called $STU$ model studied in \cite{STUkallosh}
and based on the solvable subalgebra:
\begin{equation}
Solv \left( \frac{SL(2,\IR)^3}{U(1)^3} \right) \, \subset \, Solv_3
\label{rilevanti}
\end{equation}
\par
In other words the only truely indipendent degrees of freedom of the
black hole solution are given by three complex scalar fields,
$S,T,U$.
The real parts of these scalar fields correspond to the three Cartan
generators of $Solv_3$ and have the physical interpretation of radii
of the torus compactification from $D=10$ to $D=4$. The imaginary
parts of these complex fields are generalised theta angles.
\section{Uniqueness of the N=8 abelian gauging}
Going back to the original motivations explained in the introduction,
in \cite{gaugenoi} we have studied the possible abelian gaugings of
$N=8$ supergravity. Using the solvable Lie algebra approach we have
shown that there exists a unique abelian subalgebra ${\cal A} \, \subset
 SL(8,R)\, \subset\, E_{7(7)}$ that can be gauged and this is the
 algebra $CSO(1,7)$ alreday found by Hull \cite{Hull} in his study of
 non--compact $N=8$ gaugings. The number of generators in ${\cal A}$
 is seven. The scalar potential associated with this gauging has not
 been studied in full detail and it is now our plan to study its
 properties in full generality bu using the solvable Lie algebra
 approach.

\end{document}